\documentclass[twocolumn,floatfix,aps,prl,showpacs]{revtex4-1}
\usepackage{graphicx,enumerate}

\begin{document}
\title{Finding a unifying motif of intermolecular cooperativity in protein associations}
\author{Sebasti\'an R. Accordino$^\dag$}
\author{J. Ariel Rodriguez-Fris$^\dag$}
\author{Gustavo A. Appignanesi$^\dag$}
\author{Ariel Fern\'andez$^{\ddag \clubsuit}$}
 \affiliation{
$^\dag$Secci\'on Fisicoqu\'{i}mica, INQUISUR-UNS-CONICET-Departamento de Qu\'{i}mica, Universidad Nacional del Sur, Avda. Alem 153,
8000 Bah\'{i}a Blanca, Argentina.\\
$^\ddag$Instituto Argentino de Matem\'atica ``Alberto P. Calder\'on'', CONICET, Buenos Aires 1083, Argentina.\\
$^\clubsuit$Department of Computer Science, The University of Chicago, Chicago, IL 60637}

\date{\today}

\keywords{@@@}

\begin{abstract}
At the molecular level, most biological processes entail protein associations which in turn rely on a small fraction of interfacial residues called hot spots. Here we show that hot spots share a unifying molecular attribute: they provide a third-body contribution to intermolecular cooperativity. Such motif, based on the wrapping of interfacial electrostatic interactions, is essential to maintain the integrity of the interface and can be exploited in rational drug design since such regions may serve as blueprints to engineer small molecules disruptive of protein-protein interfaces.
\end{abstract}


\maketitle

Protein associations are basic molecular processes in biology \cite{1,4,5,6,7,9,10,11,13,14,15,16,17}. In spite of their importance, their biophysical underpinnings remain a subject of debate \cite{1,4,5,6,7,9,10,11,13,14,15,16,17}. A challenging standing problem involves the characterization of hot spots \cite{1,4,5,6,7,9,10,11,13,14,15,16}. These are few in number and provide the most significant contribution to the stability of the protein-protein interface. Knowledge-based and first-principle docking potentials have been relatively successful at predicting these singular sites \cite{1,4,5,6,7,9,10,11,13,14,15,16}, fitting the outcome of probes for experimental identification such as site-directed mutation or alanine-scanning \cite{1}. These techniques assess the impact on binding free energy of the truncation of an individual residue side chain at the $\beta$-carbon. Notwithstanding these predictive successes, the physical nature of hot spots remains elusive. Even the establishment of general rules for hot-spot characterization has proven unfeasible so far, as has been explicitly recognized \cite{1,6,7} and constitutes the scope of this letter. Attempts at rationalizing the stability of protein-protein interfaces based on pairwise interactions between the two chains is inconclusive at best, as demonstrated in this letter. This leads us to focus our attention on higher order energetic contributions as a theoretical framework to explain and predict binding hot spots.   
Given the relative abundance of hydrophilic residues on the protein surface, protein associations are always confronted with the disruptive effect of polar hydration \cite{19,20}. Thus, the integrity of the protein-protein interface becomes extremely reliant on intermolecular cooperativity \cite{19,20}. We make this concept precise by invoking three-body correlations, whereupon a third nonpolar body protects an electrostatic interaction pairing the other two by contributing to the exclusion of surrounding water. Since these three-body correlations must engage the two protein molecules, the correlations must be subject to an additional constraint: One body belongs to a protein chain and the other two to its binding partner. To complete this description it is necessary to classify pairwise electrostatic interactions in terms of an abundance distribution $P(\rho)$, where $\rho$ is the number of three-body correlations associated with an interaction. This distribution is defined by its mean value $\langle \rho \rangle = \sum_{\rho} [\rho \cdot P(\rho)]$ and dispersion $\sigma = (\langle(\rho - \langle \rho \rangle)^2\rangle)^{1/2}$, which leads us to single out an underprotected interaction (UPI) as one in the tail of the distribution, that is, with $\rho \le \langle \rho \rangle - \sigma$. The UPIs are crucial in defining protein associations due to their sensitivity to critical changes in intermolecular cooperativity brought about by site-directed aminoacid substitution. As demonstrated previously \cite{19,20,22,23}, UPIs are also adhesive, hence promoters of protein association because their inherent stability increases upon approach of a third-body nonpolar group that enhances its dehydration, and de-screens the partial charges. This physical picture leads us to assert that intermolecular cooperativity will be most sensitive to site-directed mutation in two particular instances: a) When a site mutation changes the wrapping value $\rho$ of an intermolecular or intramolecular interaction decreasing it to a value below the mean $\rho$; b) When in a free protein subunit the alanine substitution raises the wrapping of a UPI and, additionally, this interaction is intermolecularly wrapped in the complex.
This analysis leads us to characterize hot spots as the residues whose alanine substitution most drastically affects intermolecular cooperativity. This conjecture is validated in this work by combinatorially dissecting the protein-protein interfaces of structurally reported complexes that have been independently studied by alanine scanning through experimental means. The analysis boils down to a decomposition of the interface into a web of three-body cooperative interactions.
Besides its scientific interest, the knowledge gained from our approach may significantly impact drug discovery endeavors \cite{24}, especially since hot spots are expected to constitute the blueprint for the design of small molecule drugs disruptive of protein-protein associations.

\begin{figure}
\includegraphics[width=0.7\linewidth]{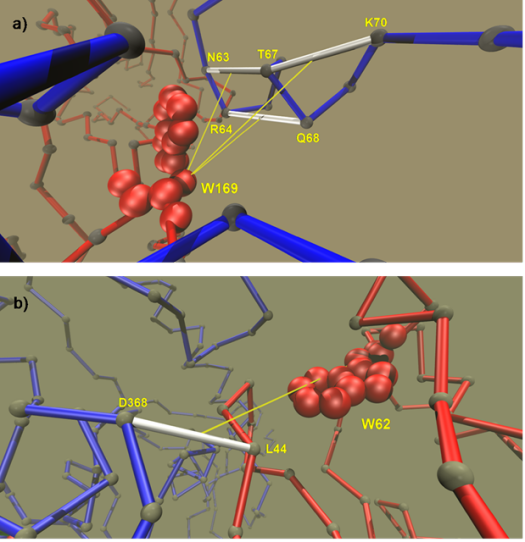}
\caption{Illustration of intermolecular cooperativity represented by three-body correlations:
a) Trp 169 (full atomic detail) of hGHbp (red chain) wrapping three intramolecular BHBs of the hGH chain (blue chain). The BHBs of the hGH chain are indicated by white sticks between the corresponding $\alpha-$carbons;
b) Similar to a) but for the complex between the HIV glycoprotein gp120 and the CD4 receptor. Here a Trp residue of the CD4 chain wraps an intermolecular BHB.}
\label{fig1}
\end{figure}

UPIs that involve hydrogen bonds (HBs) are named dehydrons. This structural motif has been extensively discussed in the literature and identified in soluble proteins with PDB-reported structure \cite{17,19,20,22,23}. Thus, the extent of hydrogen-bond protection can be determined directly from atomic coordinates. This parameter indicates the number of three-body correlations engaging the HB and is also known as the wrapping of the bond and denoted $\rho$. It is given by the number of side-chain carbonaceous nonpolar groups (CH$_n$, $n=0$, 1, 2, 3, where the carbon atom of these groups is not bonded to an electrophilic atom or polarized group) contained within a desolvation domain around the HB. Each wrapping nonpolar group represents the third body within a three-body correlation involving the HB. This domain is typically defined as the reunion of two intersecting spheres of fixed radius (~thickness of three water layers) centered at the $\alpha-$carbons of the residues paired by the hydrogen bond. In structures of PDB-reported soluble proteins, backbone hydrogen bonds (BHB) are protected on average by $\rho = 26.6 \pm 7.5$ side-chain nonpolar groups for a desolvation sphere of radius 6 \AA\ \cite{22}. Thus, structural deficiencies lie in the tail of the $\rho-$distribution, i.e. their microenvironment contains 19 or fewer nonpolar groups, so their $\rho-$value is below the mean (=26.6) minus one standard deviation (=7.5). While the statistics on $\rho-$values for backbone hydrogen bonds vary with the radius, the tails of the distribution remain invariant, thus enabling a robust identification of structural deficiencies \cite{19,20,22,23}. In the present work we are dealing with protein complexes and accordingly we compute the $\rho-$values arising from intra and inter-molecular correlations. Additionally, we consider both intramolecular and (less frequent) intermolecular BHBs. The algorithm to identify dehydrons, named ``Dehydron Calculator'', is freely accessible from the Web at the following location:
http://www.owlnet.rice.edu/~arifer/courses/DehydronCalculator.exe
The wrapping concept may be spatially represented as shown in Fig. \ref{fig1}, where two different types of three-body correlations are illustrated. Figure \ref{fig1}a) shows an instance of intermolecular wrapping of an intramolecular HB, while Fig. \ref{fig1}b) shows the wrapping of an intermolecular HB.

Our virtual alanine-scanning procedure is performed by computationally replacing each residue of a protein chain (one at a time) with alanine within the 3D structure of the complex and assessing the impact of the substitution on intermolecular cooperativity. For most residues (those with a side chain larger than that of alanine) this means truncating the residue side chain at the $\beta-$carbon so that the whole side chain is replaced by a methyl group, thus significantly reducing the extent of wrapping involving the residue. In the special case of glycine (which lacks a $\beta-$carbon) we include a methyl at the corresponding position, increasing the extent of wrapping enabled by the residue. 
The {\it in silico} scanning process entails computing the change in $\rho-$value generated by each Ala-substitution on each intra and intermolecular BHBs of the complex. In a first stage, we calculate the $\rho$ value for all BHBs from the complex structure, producing a set of wild-type $\rho-$values. For each mutated residue we perform the corresponding ala-substitution leaving all other coordinates unchanged and we recalculate the full set of $\rho-$values (mutated $\rho-$values). Then, in accord with our premise of intermolecular cooperativity, hot spots are predicted taking into account their role as intermolecular wrappers according to the following classes: a) The Ala-substitution of a residue on one chain lowers the $\rho-$value of a BHB (an intramolecular BHB in the partner protein or an intermolecular BHB) and the mutated $\rho-$value of this BHB falls below $\langle \rho \rangle$. These predicted hot spots will be labeled class A hot spots. In the cases where the final $\rho-$value falls below the dehydron threshold, $\rho=19$, (dehydron creation) these A-class hot spots will be labeled A*;
b) Alanine substitution increases the wrapping capability of a non-wrapper residue (glycine, serine, cysteine, aspartic acid or asparagine) located within the desolvation environment of a BHB of its own protein chain. In addition, the {\it intramolecular} wrapping value of the BHB is $\rho \le 19$ and this BHB is {\it intermolecularly} wrapped within the complex. These alanine substitutions raise the intramolecular $\rho$ value in $\Delta \rho = + 1$. These alterations lower the need for intermolecular wrapping upon association and the resulting predictions will be labeled class B hot spots. In the cases when the intramolecular wrapping value of the BHB is exactly $\rho=19$, we will denote these B*-class hot spots. This sub-class implies that the ala-substitution is indicative of a net intramolecular removal of a dehydron.
We decided to leave aside side chain - side chain hydrogen bonds from the cooperativity analysis based on the following grounds: The fluctuational nature of surface side chains imposes an entropic cost associated with HB formation which makes the latter marginally stable at best \cite{17}. Also, the wrapping statistics for side chain HBs are essentially flat with no clear distinction of the tails of the distribution do to the conformational richness of the side chains. An {\it a posteriori} justification for the exclusion arises from the very artifactual nature of surface side-chain HBs. Particularly misleading are the large B-factors of solvent-exposed side chains and the large hydration demands of exposed polar groups, which hinder HB formation. These artifacts would yield an overwhelming number of false positives in the cooperativity analysis of the protein-protein interface (most interfacial residues would be hot spots).
In turn, we shall not take into account salt bridges in our analysis, since they are not expected to significantly stabilize protein structure. These bridges are destabilizing with respect to hydrophobic replacement of both charged partners and charge burial has been shown to be usually destabilizing (\cite{31} and references therein). However, it is also known that for a pair of complimentary buried charges it is preferable for them to be paired by a salt bridge than to be buried isolated from each other \cite{31}. Thus, an Ala-mutation of a residue engaged in a salt bridge with its complex partner protein would be destabilizing. This trivial type of hot spots accounts for approximately 15 \% of all the hot spots in the complexes considered and obviously lies outside the scope of our cooperativity-based analysis.


We performed a cooperativity-based alanine scanning analysis on several protein-protein interfaces from complexes with PDB reported structure for which experimental alanine scanning results are available \cite{4,5} (in each case, the first protein of the complex indicated is the one mutated and we provide the PDB entry of the complex and reference of the experimental alanine scanning results): Human growth hormone receptor/Human growth hormone\cite{1} (3HHR), Trypsin inhibitor/Beta-Trypsin\cite{25} (2PTC), P53/MDM2\cite{26} (1YCR), CD4/GP120\cite{27} (1GC1), Ribonuclease inhibitor/Ribonuclease A\cite{28} (1DFJ), Colicin E9 immunity protein/Colicin E9 DNase domain\cite{29} (1BXI), Barnase/Barstar\cite{30} (1BRS), Barstar/Barnase\cite{30} (1BRS), Ribonuclease inhibitor/Angiogenin\cite{28} (14Y).

\begin{figure}
\includegraphics[width=0.9\linewidth]{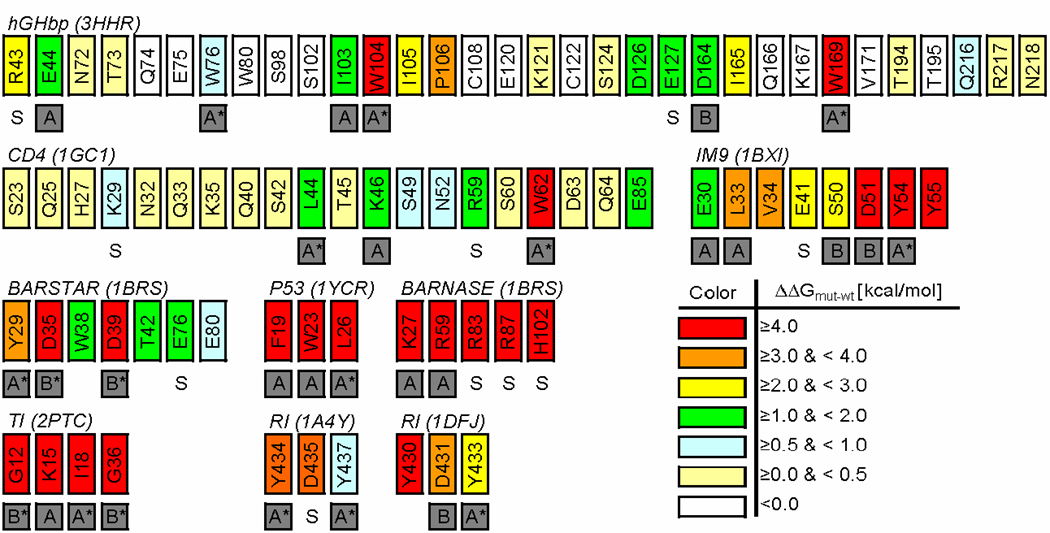}
\caption{Experimental alanine scaning probes contrasted against cooperativity-based {\it in silico} scanning for the complexes indicated.
For each case we display the portion of the protein chain or the set of residues with experimental data. The colors indicate the experimentally determined $\Delta \Delta G$ values for the corresponding hot spots, as shown in the scale at the right. The gray squares indicate our computational predictions, and the letter code is explained in the text.
}
\label{fig2}
\end{figure}

Figure \ref{fig2} displays our predictions. The experimental alanine substitution of a native protein subunit yields a change in its binding free energy $(\Delta G)$ which is denoted by $\Delta \Delta G = \Delta G_{\rm mut} - \Delta G_{\rm wt}$, (mut=mutated, wt=wild type) and is indicated with a color scale. The cooperativity-based hot-spot predictions of our method are indicated with gray squares below the corresponding residues and are denoted by A, A*, B and B*.The letter ``S'' labels trival salt bridge hot spots which are removed from the list of experimental hot spots used for the comparison with our computational method.

To quantify the predicting ability of our method, in Table \ref{table1} we show our global predictions over the whole set of protein complexes studied.

\begin{table}
\caption{Predictions obtained for the different protein complexes studied (see Table \ref{table1}).
}
\label{table1}
\begin{tabular}{|l|l|l|l|}
\hline
Experimental hot spots & \multicolumn{3}{l|}{Prediction success (percentages)} \\ \cline{2-4}
($\Delta \Delta G$ value) & A+A*+B+B* & A+A* & A*+B* \\ \hline
$ \ge 4$ kcal/mol & 89 & 61 & 56 \\ \hline
$ \ge 3$ kcal/mol & 83 & 58 & 50 \\ \hline
$ \ge 2$ kcal/mol & 79 & 54 & 46 \\ \hline
$ \ge 1$ kcal/mol & 74 & 53 & 37 \\ \hline
\end{tabular}
\end{table}

This comparison between theory and experiment reveals that our computational procedure locates most of the experimental alanine-scanning hot spots, with optimal performance (89 \% prediction success) for the most significant contributors determined experimentally ($\Delta \Delta G \ge 4$ kcal/mol). The greatest contribution to such percentage, 61 \%, corresponds to class A mutations (A and A*), while class B (B and B*) provides the remaining 28 \%. The last column of the table indicates the predictions when considering only dehydron creation, A*, and dehydron removal, B*. In consonance with our cooperativity premises, these cases are expected to constitute very important mutations and this is in fact the case, since such mutations account for 56 \% of the highly energetic mutations determined experimentally ($\Delta \Delta G \ge 4$ kcal/mol). Additionally, the wild-type $\rho-$values averaged over the residues wrapped in class A hot spots yields $\rho = 20.3$, a value higher than the dehydron threshold $(\rho = 19)$. However, when we average the mutated $\rho-$values we get a final $\rho=18$, that is, below the dehydron threshold. Thus, the dehydron threshold is in fact statistically framed by the averaged wild-type and mutated $\rho-$values for A-class hot spots, thus revealing the relevance of the qualitative wrapping differences for protein affinity.
At this point it is worth recalling that our method disregards two-body terms unless they are engaged in a three-body correlation. This approach seems natural in view of the fact that no protein-protein interface has proven trivial at the conventional pairwise level analysis \cite{1,4,5,6,7,9,10,11,13,14,15,16} and given the absence of clear rules for hot-spot prediction \cite{1,4,5,6,7,9,10,11,13,14,15,16}. This last point also makes difficult to establish a control for our results, but we have nonetheless defined an elementary one based on polar and hydrophobic complementarities. To this end, we have simply characterized residues as hydrophobic (nonpolar aromatic or aliphatic side chains) or polar (polar or charged side chains) and built a contact matrix for the complex interface. For each residue we calculated the minimum distance between its $\alpha-$carbon and the $\alpha-$carbons of the residues of the partner protein and between the centroid of its side chain and those of the partner side chains. An intermolecular contact was considered to occur when this minimum distance was below 6 \AA\ (the results are robust to moderate changes in the contact parameter and fit a criteria previously adopted for protein-protein interfaces\cite{1}). 
We applied this analysis to the hGH/hGHbp complex interface which yielded a significant level of mismatches (around 37 \%), thus indicating that the protein association cannot be simply rationalized as a search for pairwise polar-polar and hydrophobic-hydrophobic complementarity. More interestingly, when we restrict the analysis to the experimentally determined hot spots, the percentage of mismatches is slightly higher (42 \%). And if we look at the two most important hotspots (Trp 104 and Trp 169), these residues are involved in 8 mismatches and only 1 hydrophobic-hydrophobic contact. This level of mismatching seems unavoidable given the high polar content at the protein surface which becomes buried upon creation of the complex. However, when we focus on three-body interactions, we discover that many hydrophobic residues at the complex interface approach polar residues in order to wrap BHBs in which the latter are involved. 

To summarize, this letter has shown that protein-protein interfaces elude standard physico-chemical analysis. Their rationalization in terms of pairwise complementarity along the contact region is unsatisfactory, especially when it comes to understand the role of hot spots as determinants of protein associations. Against this reality, this work unravels a seemingly overlooked simple molecular motif that proves to be ubiquitous in determining protein-protein associations. This motif is an indicator of three-body intermolecular cooperativity. In essence, such effects arise as a group in one protein chain stabilizes (wraps) a preformed hydrogen bond in the partner chain or an inter-chain hydrogen bond, so that three bodies intervene in the interaction and not all three belong to the same chain. We have shown that hot-spot predictions based solely on this molecular attribute and defined by two pure combinatorial rules based on structural analysis of protein complexes, account for most (89 \%) of the hot spots experimentally determined by alanine-scanning in a set of protein complexes. Thus, the simplicity of our method contrasts with the complexity of approaches based on full fledged potentials with explicit water (where many-body terms are subsumed in all-atom interactions). We do not deny the relevance of these predictive methods, but such avenues have not proven enlightening in terms of identifying clear molecular promoters of protein associations. By contrast, the results presented in this work fulfill such imperative and might be instrumental in the design of small molecules aimed at disrupting protein-protein interfaces by fulfilling the wrapping capabilities of hot spots. We believe that our combinatorial identification of a molecular promoter of protein associations holds promise as a guidance to rational drug design.

Financial support from ANPCyT (PME 2006-1581) and CONICET is gratefully acknowledged. 

\end{document}